\documentclass[a4paper, amsfonts, amssymb, amsmath, reprint, showkeys, nofootinbib, twoside]{revtex4-1}
\usepackage[english]{babel}
\usepackage[utf8]{inputenc}
\usepackage[colorinlistoftodos, color=green!40, prependcaption]{todonotes}
\usepackage{amsthm}
\usepackage{mathtools}
\usepackage{physics}
\usepackage{xcolor}
\usepackage{graphicx}
\usepackage[left=23mm,right=13mm,top=35mm,columnsep=15pt]{geometry} 
\usepackage{adjustbox}
\usepackage{placeins}
\usepackage[T1]{fontenc}
\usepackage{lipsum}
\usepackage{csquotes}
\usepackage{dsfont}
\usepackage[pdftex, pdftitle={Article}, pdfauthor={Author}]{hyperref} 
\hypersetup{
     colorlinks=true,
     linkcolor=teal,
     filecolor=teal,
     citecolor = teal,      
     urlcolor=teal,
     }
\usepackage[most]{tcolorbox}
\newcommand{\mc}[1]{\mathcal{#1}}
\newcommand{\pd}{\partial}
\begin{document}
\title{Single copy of the Ricci flow}

\author{Rashid Alawadhi}
    \email[Correspondence email address:]{r.alawadhi@qmul.ac.uk}
    \affiliation{Centre for Theoretical Physics, School of Physical and Chemical Sciences, Queen Mary University of London, 327 Mile End Road, London E1 4NS, United Kingdom}

\date{\today} 

\begin{abstract}
    The perturbative double copy is by now a highly established correspondence between gravity and gauge theories. Non-perturbatively, information ranging from classical solutions to topological quantities on both sides have been related to each other via the double copy correspondence. In this paper we add another result, where we show that the single copy of the Ricci flow is the Yang-Mills flow on the space of connections of a principal $\text{U}(1)$-bundle.
\end{abstract}


\maketitle
\section{Introduction} \label{sec:Intro}

The double copy is a correspondence between gravity and gauge theories where one can obtain quantities of one theory from the other, be it perturbative or non-perturbative. The double copy has found tremendous success in the perturbative regime where it was successfully applied to tree and higher loop level amplitude calculations \cite{Bern:2019prr}. In its non-perturbative form the correspondence has been applied to full solutions of Einstein equations, solution generating transformations, and topological aspects of theories. Known procedures for describing exact solutions include the so called Kerr-Schild \cite{Bini:2010hrs} and Weyl double copy forms where the latter relies on the spinorial representation \cite{Monteiro:2014cda,Alawadhi:2019urr,Alawadhi:2021uie,Alfonsi:2020lub,Luna:2018dpt,Berman:2018hwd,Godazgar:2020zbv}. More recently, a twistorial formulation of the double copy has been studied in \cite{White:2020sfn,Chacon:2021wbr,Chacon:2021lox}. See \cite{Guevara:2021yud} for a related work.

The other major character in this paper is the Ricci flow, which is said to be analogous to the heat equation for Riemannian metrics. Since it was first introduced in the mathematics literature, it has found a tremendous number of applications in the fields of geometry and topology, wherein Perelman used the Ricci flow to prove the Poincaré conjecture, then an open problem since 1904 \cite{Perelman:2006un,Perelman:2006up,Perelman:2003uq}. In the physics literature, the Ricci flow appeared in Friedan's PhD thesis \cite{Friedan:1980jm} as the beta function of the non-linear sigma model. Since then the Ricci flow has found many applications in physics including string theory, gravity, black hole thermodynamics, cosmology and more \cite{Woolgar:2007vz,Petropoulos:2010zz,DeBiasio:2020xkv,Headrick:2006ti}.

The goal of this paper is to shed some light on the relation between the Ricci flow and the Yang-Mills flow, which is analogous to the former. Whereas the Ricci flow is the flow in the space of Riemannian metrics $\mathfrak{G}$, the Yang-Mills flow is in the space of, in our case, the connections $\mc{A}$ of a $\text{U}(1)$-principal bundle. Using the Kerr-Schild double copy, we obtain the Yang-Mills flow by single copying its gravitational cousin, the Ricci flow. This further solidifies the ability of the Kerr-Schild double copy in obtaining non-perturbative and global information. The relation with the RG flow of the non-linear sigma model for both the closed and open string will also be discussed.

In section 2 we shall review the basics of the Kerr-Schild single copy. For a more complete review see \cite{Monteiro:2014cda}. In section 3 we review the basics of the Ricci flow and its range of applicability. For a comprehensive review see \cite{Streets:2007kct,topping_2006}. Section 4 will deal with the main point of the paper, the single copy of the Ricci flow. In section 5 we discuss the results and their future applications.  
\section{Kerr-Schild and the single copy} \label{sec:KSSC}
In this paper we shall make use of the Kerr-Schild formulation of the classical double copy as outlined in \cite{Monteiro:2014cda}. To that end, we start with a d-dimensional Lorentzian manifold equipped with a metric $(\mc M, g)$. We require our manifold $\mc M$ to admit a Kerr-Schild metric, i.e., there exists a set of coordinates such that the metric tensor takes a particular form which we will shortly see. This form of the metric tensor exactly linearises the Einstein equations without assuming any small parameter expansion. Now we demonstrate the above with explicit formulae.

A solution belongs to the Kerr-Schild class if its metric can be written in the form
\begin{equation}
  g_{\mu\nu}=\eta_{\mu\nu}+\phi\, k_\mu k_{\nu}\, ,
\end{equation}
where we call $\eta = \text{diag}(-1,1,\ldots,1)$ the `background' metric, $\phi$ is a scalar field, $\phi:\mc{M}\rightarrow \mathbb{R}$ and $k_{\mu}$ is a co-vector, $k: T\mc{M}\rightarrow \mathbb{R}$ satisfying
\begin{equation}\label{KSID}
  \begin{split}
      k^\mu g_{\mu\nu} k^\nu &= k^\mu \eta_{\mu\nu} k^\nu = 0,\\
      k\cdot \pd k^\mu &= 0,\\
      k^{\mu}\pd_\nu k_{\mu} &= 0.
    \end{split}
  \end{equation}
i.e., it is null with respect to both the full and background metric. This has been generalised to curved background metrics. But for the purposes of this paper we shall set the background metric to be Minkowski. The inverse metric then takes the form
\begin{equation}
  g^{\mu\nu}=\eta^{\mu\nu}-\phi\, k^\mu k^{\nu} \, .
\end{equation}
 Note that the above expression is exact and no small parameter expansion has been done. In terms of the scalar field $\phi$ and co-vector $k_\mu$, the Ricci tensor and Ricci scalar are given by
\begin{equation}\label{RicciKS}
  \begin{split}
    &R^\mu{}_\nu=\frac{1}{2}(\partial^\mu\partial_\alpha(\phi\, k^\alpha k_\nu)+\partial_\nu\partial^\alpha(\phi\, k_\alpha k^\mu)-\partial^2(\phi\, k^\mu k_\nu)) \, ,\\
    &R=\partial_\mu\partial_\nu(\phi\, k^\mu k^\nu)\, ,
\end{split}
\end{equation}
where $\partial^\mu=\eta^{\mu\nu}\partial_\nu$. In the stationary spacetime case ($\partial_0 \phi=\partial_0 k^\mu=0$) one may take the time component of the Kerr-Schild vector as $k^0=1$ \cite{Monteiro:2014cda}, with the dynamics in the time component contained in $\phi$. As a consequence, the components of the Ricci tensor simplify to
\begin{equation}
  R^{0}{}_{0}=\frac{1}{2}\pd^{2}\phi\, ,
\end{equation}
\begin{equation}
  R^{i}{}_{0}=-\frac{1}{2}\partial_{j}[\partial^{i}(\phi\, k^j)-\partial^j(\phi\, k^i)]\, ,
\end{equation}
\begin{equation}
  R^{i}{}_{j}=\frac{1}{2}\partial_{l}[\partial^{i}(\phi\, k^{l}k_{j})+\partial_{j}(\phi\, k^{l}k^{i})-\partial^{l}(\phi\, k^{i}k_{j})]\, ,
\end{equation}
\begin{equation}
  R=\partial_{i}\partial_{j}(\phi\, k^{i}k^{j})\, ,
\end{equation}
where Latin indices indicate the spatial components.

Now  define a gauge field $A_{\mu}=\phi\, k_\mu$, with the Maxwell field strength $F_{\mu\nu}=\partial_{\mu} A_{\nu}-\partial_{\mu}A_\nu$. Taking the stationary case of the vacuum Einstein equations $R^{\mu}{}_{\nu}=0$, one finds that the gauge field satisfies the Abelian Maxwell equations
\begin{equation}
  \partial_{\mu}F^{\mu\nu}=\partial_{\mu}(\partial^{\mu}(\phi\, k^{\nu})-\partial^{\nu}(\phi\, k^{\mu}))=0 \, .
\end{equation}

The field strength $F_{\mu\nu}$ can be thought of the curvature of a $\text{U}(1)$-principal bundle with a Lie algebra valued connection $A\in \Omega^1(\mc U \subset\mc M, \mathfrak{u}(1))$ defined locally on a subset $\mc U$ of $\mc M$. The basic statement of the double/single copy we shall be applying is:
If $g_{\mu\nu}=\eta_{\mu\nu}+\phi\, k_\mu k_{\nu}$ is a stationary solution of Einstein's equations, then $A_{\mu}=\phi\, k_\mu$ is a solution of the Abelian Yang-Mills equations.

For the benefit of the unacquainted reader, we give a canonical example of an application of the single copy. We apply the steps mentioned above for the Schwarzschild spacetime, a spherically symmetric solution of the vacuum Einstein equations, and therefore by Birkoff's theorem, is static. We will see soon that it single copies to the Coulomb charge which is also spherically symmetric and static. In Kerr-Schild coordinates $(t,x,y,z)$ the Schwarzschild metric components have the form
\begin{equation}
  g_{\mu\nu} = \eta_{\mu\nu} + \phi\, k_{\mu} k_{\nu},
\end{equation}
with the scalar field $\phi$ and Kerr-Schild co-vector $k_{\mu}$ given by
\begin{equation}
  \phi = \frac{2M}{r},\quad k_{\mu}=\Big(-1,\frac{x^i}{r}\Big),
\end{equation}
with $M$ a constant and $r = \sqrt{x^ix_i}\in (2M,\infty)$. We then can construct the Yang-Mills connection associated with our gravitational solution by defining
\begin{equation}
  A_{\mu} = \phi\, k_{\mu} = \frac{2M}{r}\Big(-1,\frac{x_i}{r}\Big).
\end{equation}
To put the connection in a more physically suggestive form, we simply do a gauge transformation to get rid of the spatial components, upon doing so we obtain
\begin{equation}
  A = \frac{2M}{r}dt,
\end{equation}
which is the Yang-Mills field for a coulomb charge with field strength
\begin{equation}
  F = \frac{2M}{r^2}dt \wedge dr.
\end{equation}
The Coulomb charge is the unique spherically symmetric solution of Maxwell's equations in analogy with the Schwarzschild solution of Einstein equations. For a simple proof of Birkoff's theorem in the context of electrodynamics see \cite{doi:10.1119/1.13934}.

The above single copy procedure has been used in many physically relevant and interesting geometries including Kerr, Taub-NUT \cite{Luna:2015paa}, Eguchi-Hanson \cite{Berman:2018hwd}, dS and AdS spacetimes \cite{Bah:2019sda,Carrillo-Gonzalez:2017iyj}. For a double field theoretic generalisation of the Coulomb case see \cite{Kim:2019jwm}.  In this section we showed how to obtain full and non-perturbative solutions to Yang-Mills equations via single copying gravitational solutions. This is an example of a local non-perturbative map. It has also been shown that non-perturbative information can be probed and sometimes carry over from both side of the theories \cite{Alfonsi:2020lub,Alawadhi:2021uie}. In the next section we give a brief introduction to the Ricci flow.
\section{Ricci flow} 
    \label{sec:RicciFlow}

The Ricci flow is a particular PDE for the metric tensor, where the latter evolves according to its Ricci tensor. Given a $d$-dimensional Riemannian manifold endowed with a metric $(\mc M, g)$, the Ricci flow equation is 
\begin{equation}\label{RicciFloweq1}
    \frac{\pd}{\pd \lambda}g_{\mu\nu} = -2 R_{\mu\nu},\quad g_{\mu\nu}(\lambda = 0) = g^0_{\mu\nu},
\end{equation}
where both $g_{\mu\nu}$ and $R_{\mu\nu}$ are functions of the flow parameter $\lambda\in\mathbb{R}$. Eq.~\eqref{RicciFloweq1} describes the evolution of $g_{\mu\nu}(\lambda)$ in the space of metrics $\mathfrak{G}$. To illustrate this, let us see how a generic Einstein metric evolves. For Einstein spaces the Ricci tensor takes a particularly simple form given by $R_{\mu\nu} = k g_{\mu\nu}$ for some constant $k\in\mathbb{R}$. Plugging this form of the Ricci tensor into eq.~\eqref{RicciFloweq1} we find
\begin{equation}
    \frac{\pd}{\pd\lambda}g_{\mu\nu} = -2kg_{\mu\nu}.
\end{equation}
Let us take the $d$-sphere and see how it evolves under the flow. In this case we have $g_{\mu\nu}=r^2g^{S^d}_{\mu\nu}$ and $k = \frac{d-1}{r^2}$. Plugging into the the above equation we find
\begin{equation}
    r(\lambda) = \sqrt{r^2_0-2(d-1)\lambda},
\end{equation}
where $r(\lambda)$ is the radius of the $d$-sphere at a given value of the flow parameter $\lambda$. Therefore, we see that the sphere shrinks as the metric flows until it becomes singular in finite flow time given by $\lambda^* = r^2_0/(2(d-1))$. Another physically relevant example are dS and AdS spaces. In this case we have 
\begin{equation}
    R_{\mu\nu} = \frac{2\Lambda}{d-2}g^0_{\mu\nu},
\end{equation}
 and the flow gives us
 \begin{equation}
     g_{\mu\nu}(\lambda)=\bigg( 1-\frac{4\Lambda}{d-2}\lambda \bigg)g^0_{\mu\nu}.
 \end{equation}
 For $\Lambda > 0 $ the space shrinks to a point in finite flow time given by $\lambda^* = \frac{d-2}{4\Lambda}$. The case $\Lambda = 0$ is a critical point on the flow and therefore the space does not evolve. For $\Lambda<0$ the flow is immortal and the space expands indefinitely. See \cite{DeBiasio:2020xkv} for some examples and in more detail.

 As mentioned in the introduction, the Ricci flow also has relevance in the physics literature, first appearing in the context of the RG flow of the non-linear sigma model \cite{Friedan:1980jm}. Let us briefly illustrate how.

 Consider a map $f$ between two Riemannian manifolds with metrics $\gamma$ and $g$
\begin{equation}\label{mapf}
    f: (\mc B, \gamma)\rightarrow (\mc M, g),
\end{equation}
where $\mc B$ is 2-dimensional and $\mc M$ is $d$-dimensional. An action integral is given by
\begin{equation}
    S[f; \gamma, g] =\frac{1}{4\pi\alpha'} \int_{\mc V \subset \mc B}(\gamma, f^{*}g)_{\gamma}\omega_\gamma.
\end{equation}
where $\alpha'\in\mathbb{R}$ is some constant, $f^* g$ is the pullback of the metric $g$ of $\mc M$ to $\mc B$. The parenthesis $({\,},{\,})_{\gamma}$ denotes contraction with metric $\gamma_{ab}$. Finally, $\omega_\gamma = \sqrt{|\gamma|}d^2x$ denotes the volume-form on $\mc B$. Therefore, our action can be written more explicitly as

\begin{equation}
    S[f; \gamma, g] = \frac{1}{4\pi\alpha'}\int_{\mc V}\gamma^{ab}\frac{\pd y^\mu}{\pd x^a}\frac{\pd y^\nu}{\pd x^b} g_{\mu\nu}\sqrt{|\gamma|}d^2x ,
\end{equation}

where $x^a=(x^1,x^2)$ are coordinates on $\mc V\subset \mc B$ and $y^\mu=(y^1,\ldots,y^d)$ on $\mc M$. In the context of string theory one would call $\mc B$ the world-sheet, $\mc M$ spacetime, and $\alpha' \in \mathbb{R}$ the Regge slope. Writing the action for a 2D world-sheet in conformal gauge \cite{Green:2012oqa} where $\gamma_{\mu\nu}\rightarrow\eta_{\mu\nu}$ we get
\begin{equation}
      S[f; \gamma, g] = \frac{1}{4\pi\alpha'}\int_{\mc V} \pd_a y^\mu \pd^{a} y^\nu g_{\mu\nu} d^2x,
\end{equation}
where a contraction with the flat world-sheet metric is done. The non-linear sigma model action describes the propagation of a closed string on a $d$-dimensional spacetime $\mc (M,g)$. Now if we view the target space metric $g_{ab}$ as a coupling constant governing the strength of the interaction, expand around a classical background $y^\mu(x) = \bar{y}^\mu + \sqrt{\alpha'}Y^\mu (x)$, and compute the amplitude at one loop order we find that we have to renormalise to avoid divergences. The renormalisation results in a correction to the target space metric given by $g_{\mu\nu} \rightarrow g_{\mu\nu} + \alpha' R_{\mu\nu} $, where $R_{\mu\nu}$ is the Ricci tensor. The beta function is indeed given by
\begin{equation}\label{betaG}
    \beta_{\mu\nu}^g \coloneqq \frac{\pd}{\pd \lambda}g_{\mu\nu} = \alpha' R_{\mu\nu}
\end{equation}
for some energy scale $\lambda \in \mathbb{R}$. Indeed, eq.~\eqref{betaG} is the Ricci flow equation(with $\alpha' = -2 $).

In this section we reviewed the basics of the Ricci flow with some examples. We also gave a brief description of its relevance in the physics literature in the context of string theory. In the following section we introduce the main result of the paper, the Yang-Mills flow as the single copy of the Ricci flow. Physically, the Yang-Mills flow is known to be the first order correction to the Yang-Mills field of an open string.
\section{Yang-Mills flow}
\label{sec:YMFlow}
To obtain the single copy of the Ricci flow of eq.~\eqref{RicciFloweq1}, and in accordance with our method of the double copy, we restrict our class of metrics to the algebraically special Kerr-Schild metrics, thus writing the Ricci flow equations as
\begin{equation}
    \frac{\pd}{\pd\lambda}\Big( \eta_{\mu\nu} + \phi(\lambda) k_{\mu}(\lambda)k_{\nu}(\lambda) \Big) = -2 R_{\mu\nu},
\end{equation}

Note that the first term on the LHS does not depend on the flow parameter $\lambda$, hence $\pd_\lambda \eta_{\mu\nu} = 0$. Dropping the explicit dependence on $\lambda$ we now have
\begin{equation}
    \frac{\pd}{\pd\lambda}( \phi\, k_{\mu}k_{\nu} ) = -2 R_{\mu\nu}.
\end{equation}

Now the Ricci tensor and hence the vacuum Einstein equations only linearise in $\phi$ if one of the indices is contravariant \cite{Monteiro:2014cda}. Therefore we raise one index by contracting it with $g^{\mu\alpha}$, thus obtaining

\begin{equation}
    \frac{\pd}{\pd\lambda}(\phi\, k^{\alpha}k_{\nu})-\phi^2 k^{\alpha}k_{\mu}\frac{\pd k^{\mu}}{\pd\lambda}= -2 R^{\alpha}{}_{\nu},
\end{equation}
where we used the fact that $k_{\mu}$ is null as given by eq.~\eqref{KSID}. In fact, the second term on the LHS is zero since

\begin{equation}
   k_{\mu} \frac{\pd k^{\mu}}{\pd\lambda} = \frac{1}{2}\frac{\pd}{\pd\lambda}(k_\mu k^\mu) = 0.
\end{equation}
Therefore, the Ricci flow equation takes on a particularly simple form ready for single copying:
\begin{equation}\label{KSmixFlow}
    \frac{\pd}{\pd\lambda}(\phi\, k^{\alpha}k_{\nu}) = -2 R^{\alpha}{}_{\nu}.
\end{equation}
Now if we simply follow the argument given under eq.~\eqref{RicciKS} and define a Yang-Mills connection form $A_{\nu}=\phi\, k_{\nu}$ with $\alpha = 0$, we find the Yang-Mills flow\footnote[1]{For a discussion on parabolicity, gauge, and diffeomorphism invariance for both the Ricci and Yang-Mills flow see \cite{Headrick:2006ti,Streets:2007kct}.}
\begin{equation}\label{YMFlow}
    \frac{\pd}{\pd\lambda}A_{\mu} = -\pd_{\nu}F_{\mu}{}^{\nu}
\end{equation}

Given our choice of applying the Kerr-Schild single copy on a flat Minkowski background, equation~\eqref{YMFlow} should be interpreted as the flow in the space of connections of a principal $\text{U}(1)$-bundle $(P,\mathbb{R}^{1,d-1},\text{U}(1))$ with curvature $F \in \Omega^{2}(\mathbb{R}^{1,d-1},\mathfrak{u}(1))$. Again, eq.~\eqref{YMFlow} is analogous to a diffusion equation for the gauge field $A_{\mu}$ since with $F_{\mu\nu} = 2\pd_{[\mu}A_{\nu]}$ we have
\begin{equation}
    \frac{\pd}{\pd \lambda}A_{\mu} = \pd^2A_{\mu} - \pd_\nu \pd_\mu A^{\nu}.
\end{equation}
Now that we have introduced the Yang-Mills flow, let us see where it appears in the physics literature.

In studying the interaction of closed and open strings in background fields \cite{Abouelsaood:1986gd,Fradkin:1985qd,Callan:1986bc}, divergences arise which upon cancelling them give rise to a set of loop corrected beta functions. One of those is the beta function for the open string coupled to the electromagnetic gauge field potential $A_{\mu}$. Fradkin and Tseytlin showed in \cite{Fradkin:1985qd} that one finds non-linear corrections to Maxwell's equations upon cancelling the divergences that appear at tree and loop level in the open string sigma model. Subsequently, it was further studied by others in \cite{Abouelsaood:1986gd,Callan:1986bc}. Let us briefly show how the above comes about. Starting from the sigma model functional for an open string coupled to a background gauge field $A_{\mu}$,
\begin{equation}
    \begin{split}\label{openStringSigma}
        S[f; \gamma, \eta]= \frac{1}{4\pi\alpha'}\int_{\mc{V} \in \mc{B}}\gamma^{a b}&\pd_{a}y^{\mu}\pd_{b}y_{\mu}\sqrt{\gamma}d^{2}x\\
    & - \oint_{\pd\mc{B}}A_{\mu}(y)\frac{\pd}{\pd s}y^{\mu}ds,
    \end{split}
\end{equation}
where the second integral is over the boundary induced by $\gamma_{ab}$. We have taken the target space to be Minkowski $\eta_{\mu\nu} = \text{diag}(-1,1,\ldots,1)$. The beta function for eq.~\eqref{openStringSigma} was exactly found for all orders in $\alpha'$ and \emph{lowest in derivatives} of the field strength $F_{\mu\nu}$(see \cite{Abouelsaood:1986gd} for a detailed derivation),
\begin{equation}\label{betaOpen}
    \beta_{\mu}^{A} \coloneqq \frac{\pd}{\pd \lambda}A_{\mu} = 2\pi\alpha' \pd^{\nu}F_{\mu}{}^{\delta}\Big[\mathds{1} - (2\pi\alpha' F)^2\Big]^{-1}_{\delta\nu},
\end{equation}
which is in fact obtainable from varying the Born-Infeld functional with respect to $A_{\mu}$ \cite{Callan:1986bc}
\begin{equation}
    S_\text{{B-I}} = \int d^dx\sqrt{\det(\mathds{1}+2\pi\alpha' F)}.
\end{equation}
For fields $F \ll 1/\alpha'$ we can expand in powers of $\alpha'$
\begin{equation}
    S_\text{{B-I}} = \int d^dx \Big(1 + \frac{(2\pi\alpha')^2}{4}F_{\mu\nu}F^{\mu\nu} + O(\alpha'^3)\Big),
\end{equation}
where the the second term in the action is just the Yang-Mills Lagrangian. Indeed, the beta function \eqref{betaOpen} admits the same expansion in $\alpha'$
\begin{equation}\label{betaYM}
    \beta_{\mu}^{A} = 2\pi\alpha' \partial_{\nu}F_{\mu}{}^{\nu} + O(F^2,\pd^2 F),
\end{equation}
 giving us one of Maxwell's equations since conformal symmetry demands the beta function to be equal to zero. Treating the beta function more generally, say off-shell, eq.~\eqref{betaYM} is nothing but the Yang-Mills flow equation for a $\text{U}(1)$-connection. It is remarkable that the single copy relates the beta function of the closed string on a curved background to the beta function of the open string on a flat background that is coupled to a gauge field. This is in agreement with the spirit of the single copy and also reminiscent of the KLT relations that relate the amplitude of the closed string to the square of the open string \cite{Kawai:1985xq}. More precisely, the single copy relates the flow in the space of metrics $\mathfrak{G}$ to the flow in the space of $\text{U}(1)$-connections $\mc A$. Indeed, the space of metrics is restricted to a particular subset of metrics that are compatible with our single copy procedure.
\section{The zeroth copy}
\label{sec:zerothFlow}
Indeed, in the context of the double copy, there is the notion of the zeroth copy where one goes from the gauge theory side to a scalar theory called the biadoint theory by applying the zeroth copy. Schematically,
\begin{equation}
    g_{\mu\nu} = \eta_{\mu\nu} + \phi\, k_{\mu}k_{\nu} \rightarrow A_{\mu} = \phi\, k_{\mu} \rightarrow \phi
\end{equation}
where the arrow indicates the stripping off of one Kerr-Schild null vector $k_{\mu}$. Furthermore, we can either take the zeroth copy of the Yang-Mills flow eq.~\eqref{YMFlow} or simply start with the Kerr-Schild Ricci flow eq.~\eqref{RicciKS} and set $\mu=\nu=0$ to obtain the flow equation for the scalar field $\phi$
\begin{equation}\label{phiFlow}
    \frac{\pd}{\pd\lambda}\phi = \pd^2 \phi,
\end{equation}
which is nothing but the familiar heat equation for $\phi(\lambda, x)$. As implied by the Kerr-Schild Einstein equations, at the critical point of eq.~\eqref{phiFlow}, the scalar field satisfies
\begin{equation}
    \pd^2\phi = 0,
\end{equation}
i.e. it is a harmonic function. The relation to the dilaton beta function of the non-linear sigma model is unclear since the zeroth copy corresponds to biajoint scalar theory \cite{Monteiro:2014cda}, which is unrelated to the dilaton appearing in the non-linear sigma model \cite{Polchinski:1998rq}.
\section{Summary and discussion}
\label{sec:Disc}
In this paper we have reviewed the non-perturbative double copy in its Kerr-Schild form and gave a simple example of a Schwarzschild black hole single copying to a point-like Coulomb charge. In section two, we have introduced the Ricci flow with some examples and its appearance in the physics literature as the first order (in $\alpha'$) correction to the target space metric tensor. In the following section the Yang-Mills flow is obtained by single copying the Ricci flow equation. Physically, the Yang-Mills flow is the first order correction to the Yang-Mills gauge field arising from the beta function of the open string theory. This has been done to all orders in $\alpha'$ where the exact form of the beta function was found to be related to the Born-Infeld non-linear generalisation of Maxwell's electromagnetism.

Two obvious avenues to extend this work into are, firstly, generalising the classical double copy relation to higher orders in the Regge slope $\alpha'$. While mathematically the map from the Ricci flow eq.~\eqref{RicciFloweq1} to the Yang-Mills flow eq.~\eqref{YMFlow} is non-perturbative, physically the flows are perturbative corrections obtained by calculating amplitudes of the non-linear sigma model. Recently, Pasarin and Tseytlin \cite{Pasarin:2020qoa} obtained a metric tensor satisfying the Einstein equations by double copying a gauge field corrected by the Born-Infeld theory, hinting at a non-linear generalisation of the classical double copy. See \cite{Easson:2020esh} for a related work involving the double copy of non-singular black holes. Secondly, the Ricci flow has been used in studying the thermodynamics and phase transitions of black holes \cite{Headrick:2006ti,DeBiasio:2020xkv}. It would be interesting to see what are the parallel behaviours and quantities on the gauge theory side by applying the single copy procedure.
\section*{Acknowledgements} \label{sec:acknowledgements}
    The author would like to thank David Berman, David Peinador Veiga and and Chris White for useful discussions and general support. The author is funded by the UAE Ministry of Education.
\bibliography{Refs.bib}

\begin{thebibliography}{36}%
\makeatletter
\providecommand \@ifxundefined [1]{%
 \@ifx{#1\undefined}
}%
\providecommand \@ifnum [1]{%
 \ifnum #1\expandafter \@firstoftwo
 \else \expandafter \@secondoftwo
 \fi
}%
\providecommand \@ifx [1]{%
 \ifx #1\expandafter \@firstoftwo
 \else \expandafter \@secondoftwo
 \fi
}%
\providecommand \natexlab [1]{#1}%
\providecommand \enquote  [1]{``#1''}%
\providecommand \bibnamefont  [1]{#1}%
\providecommand \bibfnamefont [1]{#1}%
\providecommand \citenamefont [1]{#1}%
\providecommand \href@noop [0]{\@secondoftwo}%
\providecommand \href [0]{\begingroup \@sanitize@url \@href}%
\providecommand \@href[1]{\@@startlink{#1}\@@href}%
\providecommand \@@href[1]{\endgroup#1\@@endlink}%
\providecommand \@sanitize@url [0]{\catcode `\\12\catcode `\$12\catcode
  `\&12\catcode `\#12\catcode `\^12\catcode `\_12\catcode `\%12\relax}%
\providecommand \@@startlink[1]{}%
\providecommand \@@endlink[0]{}%
\providecommand \url  [0]{\begingroup\@sanitize@url \@url }%
\providecommand \@url [1]{\endgroup\@href {#1}{\urlprefix }}%
\providecommand \urlprefix  [0]{URL }%
\providecommand \Eprint [0]{\href }%
\providecommand \doibase [0]{http://dx.doi.org/}%
\providecommand \selectlanguage [0]{\@gobble}%
\providecommand \bibinfo  [0]{\@secondoftwo}%
\providecommand \bibfield  [0]{\@secondoftwo}%
\providecommand \translation [1]{[#1]}%
\providecommand \BibitemOpen [0]{}%
\providecommand \bibitemStop [0]{}%
\providecommand \bibitemNoStop [0]{.\EOS\space}%
\providecommand \EOS [0]{\spacefactor3000\relax}%
\providecommand \BibitemShut  [1]{\csname bibitem#1\endcsname}%
\let\auto@bib@innerbib\@empty
\bibitem [{\citenamefont {Bern}\ \emph {et~al.}(2019)\citenamefont {Bern},
  \citenamefont {Carrasco}, \citenamefont {Chiodaroli}, \citenamefont
  {Johansson},\ and\ \citenamefont {Roiban}}]{Bern:2019prr}%
  \BibitemOpen
  \bibfield  {author} {\bibinfo {author} {\bibfnamefont {Z.}~\bibnamefont
  {Bern}}, \bibinfo {author} {\bibfnamefont {J.~J.}\ \bibnamefont {Carrasco}},
  \bibinfo {author} {\bibfnamefont {M.}~\bibnamefont {Chiodaroli}}, \bibinfo
  {author} {\bibfnamefont {H.}~\bibnamefont {Johansson}}, \ and\ \bibinfo
  {author} {\bibfnamefont {R.}~\bibnamefont {Roiban}},\ }\href@noop {} {\
  (\bibinfo {year} {2019})},\ \Eprint {http://arxiv.org/abs/1909.01358}
  {arXiv:1909.01358 [hep-th]} \BibitemShut {NoStop}%
\bibitem [{\citenamefont {Bini}\ \emph {et~al.}(2010)\citenamefont {Bini},
  \citenamefont {Geralico},\ and\ \citenamefont {Kerr}}]{Bini:2010hrs}%
  \BibitemOpen
  \bibfield  {author} {\bibinfo {author} {\bibfnamefont {D.}~\bibnamefont
  {Bini}}, \bibinfo {author} {\bibfnamefont {A.}~\bibnamefont {Geralico}}, \
  and\ \bibinfo {author} {\bibfnamefont {R.~P.}\ \bibnamefont {Kerr}},\ }\href
  {\doibase 10.1142/S0219887810004518} {\bibfield  {journal} {\bibinfo
  {journal} {Int. J. Geom. Meth. Mod. Phys.}\ }\textbf {\bibinfo {volume}
  {7}},\ \bibinfo {pages} {693} (\bibinfo {year} {2010})},\ \Eprint
  {http://arxiv.org/abs/1408.4601} {arXiv:1408.4601 [gr-qc]} \BibitemShut
  {NoStop}%
\bibitem [{\citenamefont {Monteiro}\ \emph {et~al.}(2014)\citenamefont
  {Monteiro}, \citenamefont {O'Connell},\ and\ \citenamefont
  {White}}]{Monteiro:2014cda}%
  \BibitemOpen
  \bibfield  {author} {\bibinfo {author} {\bibfnamefont {R.}~\bibnamefont
  {Monteiro}}, \bibinfo {author} {\bibfnamefont {D.}~\bibnamefont {O'Connell}},
  \ and\ \bibinfo {author} {\bibfnamefont {C.~D.}\ \bibnamefont {White}},\
  }\href {\doibase 10.1007/JHEP12(2014)056} {\bibfield  {journal} {\bibinfo
  {journal} {JHEP}\ }\textbf {\bibinfo {volume} {12}},\ \bibinfo {pages} {056}
  (\bibinfo {year} {2014})},\ \Eprint {http://arxiv.org/abs/1410.0239}
  {arXiv:1410.0239 [hep-th]} \BibitemShut {NoStop}%
\bibitem [{\citenamefont {Alawadhi}\ \emph {et~al.}(2020)\citenamefont
  {Alawadhi}, \citenamefont {Berman}, \citenamefont {Spence},\ and\
  \citenamefont {Peinador~Veiga}}]{Alawadhi:2019urr}%
  \BibitemOpen
  \bibfield  {author} {\bibinfo {author} {\bibfnamefont {R.}~\bibnamefont
  {Alawadhi}}, \bibinfo {author} {\bibfnamefont {D.~S.}\ \bibnamefont
  {Berman}}, \bibinfo {author} {\bibfnamefont {B.}~\bibnamefont {Spence}}, \
  and\ \bibinfo {author} {\bibfnamefont {D.}~\bibnamefont {Peinador~Veiga}},\
  }\href {\doibase 10.1007/JHEP03(2020)059} {\bibfield  {journal} {\bibinfo
  {journal} {JHEP}\ }\textbf {\bibinfo {volume} {03}},\ \bibinfo {pages} {059}
  (\bibinfo {year} {2020})},\ \Eprint {http://arxiv.org/abs/1911.06797}
  {arXiv:1911.06797 [hep-th]} \BibitemShut {NoStop}%
\bibitem [{\citenamefont {Alawadhi}\ \emph {et~al.}(2021)\citenamefont
  {Alawadhi}, \citenamefont {Berman}, \citenamefont {White},\ and\
  \citenamefont {Wikeley}}]{Alawadhi:2021uie}%
  \BibitemOpen
  \bibfield  {author} {\bibinfo {author} {\bibfnamefont {R.}~\bibnamefont
  {Alawadhi}}, \bibinfo {author} {\bibfnamefont {D.~S.}\ \bibnamefont
  {Berman}}, \bibinfo {author} {\bibfnamefont {C.~D.}\ \bibnamefont {White}}, \
  and\ \bibinfo {author} {\bibfnamefont {S.}~\bibnamefont {Wikeley}},\ }\href
  {\doibase 10.1007/JHEP10(2021)229} {\bibfield  {journal} {\bibinfo  {journal}
  {JHEP}\ }\textbf {\bibinfo {volume} {10}},\ \bibinfo {pages} {229} (\bibinfo
  {year} {2021})},\ \Eprint {http://arxiv.org/abs/2107.01114} {arXiv:2107.01114
  [hep-th]} \BibitemShut {NoStop}%
\bibitem [{\citenamefont {Alfonsi}\ \emph {et~al.}(2020)\citenamefont
  {Alfonsi}, \citenamefont {White},\ and\ \citenamefont
  {Wikeley}}]{Alfonsi:2020lub}%
  \BibitemOpen
  \bibfield  {author} {\bibinfo {author} {\bibfnamefont {L.}~\bibnamefont
  {Alfonsi}}, \bibinfo {author} {\bibfnamefont {C.~D.}\ \bibnamefont {White}},
  \ and\ \bibinfo {author} {\bibfnamefont {S.}~\bibnamefont {Wikeley}},\ }\href
  {\doibase 10.1007/JHEP07(2020)091} {\bibfield  {journal} {\bibinfo  {journal}
  {JHEP}\ }\textbf {\bibinfo {volume} {07}},\ \bibinfo {pages} {091} (\bibinfo
  {year} {2020})},\ \Eprint {http://arxiv.org/abs/2004.07181} {arXiv:2004.07181
  [hep-th]} \BibitemShut {NoStop}%
\bibitem [{\citenamefont {Luna}\ \emph {et~al.}(2019)\citenamefont {Luna},
  \citenamefont {Monteiro}, \citenamefont {Nicholson},\ and\ \citenamefont
  {O'Connell}}]{Luna:2018dpt}%
  \BibitemOpen
  \bibfield  {author} {\bibinfo {author} {\bibfnamefont {A.}~\bibnamefont
  {Luna}}, \bibinfo {author} {\bibfnamefont {R.}~\bibnamefont {Monteiro}},
  \bibinfo {author} {\bibfnamefont {I.}~\bibnamefont {Nicholson}}, \ and\
  \bibinfo {author} {\bibfnamefont {D.}~\bibnamefont {O'Connell}},\ }\href
  {\doibase 10.1088/1361-6382/ab03e6} {\bibfield  {journal} {\bibinfo
  {journal} {Class. Quant. Grav.}\ }\textbf {\bibinfo {volume} {36}},\ \bibinfo
  {pages} {065003} (\bibinfo {year} {2019})},\ \Eprint
  {http://arxiv.org/abs/1810.08183} {arXiv:1810.08183 [hep-th]} \BibitemShut
  {NoStop}%
\bibitem [{\citenamefont {Berman}\ \emph {et~al.}(2019)\citenamefont {Berman},
  \citenamefont {Chac\'on}, \citenamefont {Luna},\ and\ \citenamefont
  {White}}]{Berman:2018hwd}%
  \BibitemOpen
  \bibfield  {author} {\bibinfo {author} {\bibfnamefont {D.~S.}\ \bibnamefont
  {Berman}}, \bibinfo {author} {\bibfnamefont {E.}~\bibnamefont {Chac\'on}},
  \bibinfo {author} {\bibfnamefont {A.}~\bibnamefont {Luna}}, \ and\ \bibinfo
  {author} {\bibfnamefont {C.~D.}\ \bibnamefont {White}},\ }\href {\doibase
  10.1007/JHEP01(2019)107} {\bibfield  {journal} {\bibinfo  {journal} {JHEP}\
  }\textbf {\bibinfo {volume} {01}},\ \bibinfo {pages} {107} (\bibinfo {year}
  {2019})},\ \Eprint {http://arxiv.org/abs/1809.04063} {arXiv:1809.04063
  [hep-th]} \BibitemShut {NoStop}%
\bibitem [{\citenamefont {Godazgar}\ \emph {et~al.}(2021)\citenamefont
  {Godazgar}, \citenamefont {Godazgar}, \citenamefont {Monteiro}, \citenamefont
  {Peinador~Veiga},\ and\ \citenamefont {Pope}}]{Godazgar:2020zbv}%
  \BibitemOpen
  \bibfield  {author} {\bibinfo {author} {\bibfnamefont {H.}~\bibnamefont
  {Godazgar}}, \bibinfo {author} {\bibfnamefont {M.}~\bibnamefont {Godazgar}},
  \bibinfo {author} {\bibfnamefont {R.}~\bibnamefont {Monteiro}}, \bibinfo
  {author} {\bibfnamefont {D.}~\bibnamefont {Peinador~Veiga}}, \ and\ \bibinfo
  {author} {\bibfnamefont {C.~N.}\ \bibnamefont {Pope}},\ }\href {\doibase
  10.1103/PhysRevLett.126.101103} {\bibfield  {journal} {\bibinfo  {journal}
  {Phys. Rev. Lett.}\ }\textbf {\bibinfo {volume} {126}},\ \bibinfo {pages}
  {101103} (\bibinfo {year} {2021})},\ \Eprint
  {http://arxiv.org/abs/2010.02925} {arXiv:2010.02925 [hep-th]} \BibitemShut
  {NoStop}%
\bibitem [{\citenamefont {White}(2021)}]{White:2020sfn}%
  \BibitemOpen
  \bibfield  {author} {\bibinfo {author} {\bibfnamefont {C.~D.}\ \bibnamefont
  {White}},\ }\href {\doibase 10.1103/PhysRevLett.126.061602} {\bibfield
  {journal} {\bibinfo  {journal} {Phys. Rev. Lett.}\ }\textbf {\bibinfo
  {volume} {126}},\ \bibinfo {pages} {061602} (\bibinfo {year} {2021})},\
  \Eprint {http://arxiv.org/abs/2012.02479} {arXiv:2012.02479 [hep-th]}
  \BibitemShut {NoStop}%
\bibitem [{\citenamefont {Chac\'on}\ \emph
  {et~al.}(2021{\natexlab{a}})\citenamefont {Chac\'on}, \citenamefont {Nagy},\
  and\ \citenamefont {White}}]{Chacon:2021wbr}%
  \BibitemOpen
  \bibfield  {author} {\bibinfo {author} {\bibfnamefont {E.}~\bibnamefont
  {Chac\'on}}, \bibinfo {author} {\bibfnamefont {S.}~\bibnamefont {Nagy}}, \
  and\ \bibinfo {author} {\bibfnamefont {C.~D.}\ \bibnamefont {White}},\ }\href
  {\doibase 10.1007/JHEP05(2021)239} {\bibfield  {journal} {\bibinfo  {journal}
  {JHEP}\ }\textbf {\bibinfo {volume} {05}},\ \bibinfo {pages} {2239} (\bibinfo
  {year} {2021}{\natexlab{a}})},\ \Eprint {http://arxiv.org/abs/2103.16441}
  {arXiv:2103.16441 [hep-th]} \BibitemShut {NoStop}%
\bibitem [{\citenamefont {Chac\'on}\ \emph
  {et~al.}(2021{\natexlab{b}})\citenamefont {Chac\'on}, \citenamefont {Nagy},\
  and\ \citenamefont {White}}]{Chacon:2021lox}%
  \BibitemOpen
  \bibfield  {author} {\bibinfo {author} {\bibfnamefont {E.}~\bibnamefont
  {Chac\'on}}, \bibinfo {author} {\bibfnamefont {S.}~\bibnamefont {Nagy}}, \
  and\ \bibinfo {author} {\bibfnamefont {C.~D.}\ \bibnamefont {White}},\
  }\href@noop {} {\  (\bibinfo {year} {2021}{\natexlab{b}})},\ \Eprint
  {http://arxiv.org/abs/2112.06764} {arXiv:2112.06764 [hep-th]} \BibitemShut
  {NoStop}%
\bibitem [{\citenamefont {Guevara}(2021)}]{Guevara:2021yud}%
  \BibitemOpen
  \bibfield  {author} {\bibinfo {author} {\bibfnamefont {A.}~\bibnamefont
  {Guevara}},\ }\href@noop {} {\  (\bibinfo {year} {2021})},\ \Eprint
  {http://arxiv.org/abs/2112.05111} {arXiv:2112.05111 [hep-th]} \BibitemShut
  {NoStop}%
\bibitem [{\citenamefont {Perelman}(2006{\natexlab{a}})}]{Perelman:2006un}%
  \BibitemOpen
  \bibfield  {author} {\bibinfo {author} {\bibfnamefont {G.}~\bibnamefont
  {Perelman}},\ }\href@noop {} {\  (\bibinfo {year} {2006}{\natexlab{a}})},\
  \Eprint {http://arxiv.org/abs/math/0211159} {arXiv:math/0211159} \BibitemShut
  {NoStop}%
\bibitem [{\citenamefont {Perelman}(2006{\natexlab{b}})}]{Perelman:2006up}%
  \BibitemOpen
  \bibfield  {author} {\bibinfo {author} {\bibfnamefont {G.}~\bibnamefont
  {Perelman}},\ }\href@noop {} {\  (\bibinfo {year} {2006}{\natexlab{b}})},\
  \Eprint {http://arxiv.org/abs/math/0303109} {arXiv:math/0303109} \BibitemShut
  {NoStop}%
\bibitem [{\citenamefont {Perelman}(2003)}]{Perelman:2003uq}%
  \BibitemOpen
  \bibfield  {author} {\bibinfo {author} {\bibfnamefont {G.}~\bibnamefont
  {Perelman}},\ }\href@noop {} {\  (\bibinfo {year} {2003})},\ \Eprint
  {http://arxiv.org/abs/math/0307245} {arXiv:math/0307245} \BibitemShut
  {NoStop}%
\bibitem [{\citenamefont {Friedan}(1985)}]{Friedan:1980jm}%
  \BibitemOpen
  \bibfield  {author} {\bibinfo {author} {\bibfnamefont {D.~H.}\ \bibnamefont
  {Friedan}},\ }\href {\doibase 10.1016/0003-4916(85)90384-7} {\bibfield
  {journal} {\bibinfo  {journal} {Annals Phys.}\ }\textbf {\bibinfo {volume}
  {163}},\ \bibinfo {pages} {318} (\bibinfo {year} {1985})}\BibitemShut
  {NoStop}%
\bibitem [{\citenamefont {Woolgar}(2008)}]{Woolgar:2007vz}%
  \BibitemOpen
  \bibfield  {author} {\bibinfo {author} {\bibfnamefont {E.}~\bibnamefont
  {Woolgar}},\ }\href {\doibase 10.1139/P07-146} {\bibfield  {journal}
  {\bibinfo  {journal} {Can. J. Phys.}\ }\textbf {\bibinfo {volume} {86}},\
  \bibinfo {pages} {645} (\bibinfo {year} {2008})},\ \Eprint
  {http://arxiv.org/abs/0708.2144} {arXiv:0708.2144 [hep-th]} \BibitemShut
  {NoStop}%
\bibitem [{\citenamefont {Petropoulos}(2010)}]{Petropoulos:2010zz}%
  \BibitemOpen
  \bibfield  {author} {\bibinfo {author} {\bibfnamefont {P.~M.}\ \bibnamefont
  {Petropoulos}},\ }\href {\doibase 10.1002/prop.201000059} {\bibfield
  {journal} {\bibinfo  {journal} {Fortsch. Phys.}\ }\textbf {\bibinfo {volume}
  {58}},\ \bibinfo {pages} {839} (\bibinfo {year} {2010})},\ \Eprint
  {http://arxiv.org/abs/1011.1106} {arXiv:1011.1106 [hep-th]} \BibitemShut
  {NoStop}%
\bibitem [{\citenamefont {De~Biasio}\ and\ \citenamefont
  {L\"ust}(2020)}]{DeBiasio:2020xkv}%
  \BibitemOpen
  \bibfield  {author} {\bibinfo {author} {\bibfnamefont {D.}~\bibnamefont
  {De~Biasio}}\ and\ \bibinfo {author} {\bibfnamefont {D.}~\bibnamefont
  {L\"ust}},\ }\href {\doibase 10.1002/prop.202000053} {\bibfield  {journal}
  {\bibinfo  {journal} {Fortsch. Phys.}\ }\textbf {\bibinfo {volume} {68}},\
  \bibinfo {pages} {2000053} (\bibinfo {year} {2020})},\ \Eprint
  {http://arxiv.org/abs/2006.03076} {arXiv:2006.03076 [hep-th]} \BibitemShut
  {NoStop}%
\bibitem [{\citenamefont {Headrick}\ and\ \citenamefont
  {Wiseman}(2006)}]{Headrick:2006ti}%
  \BibitemOpen
  \bibfield  {author} {\bibinfo {author} {\bibfnamefont {M.}~\bibnamefont
  {Headrick}}\ and\ \bibinfo {author} {\bibfnamefont {T.}~\bibnamefont
  {Wiseman}},\ }\href {\doibase 10.1088/0264-9381/23/23/006} {\bibfield
  {journal} {\bibinfo  {journal} {Class. Quant. Grav.}\ }\textbf {\bibinfo
  {volume} {23}},\ \bibinfo {pages} {6683} (\bibinfo {year} {2006})},\ \Eprint
  {http://arxiv.org/abs/hep-th/0606086} {arXiv:hep-th/0606086} \BibitemShut
  {NoStop}%
\bibitem [{\citenamefont {Streets}(2007)}]{Streets:2007kct}%
  \BibitemOpen
  \bibfield  {author} {\bibinfo {author} {\bibfnamefont {J.~D.}\ \bibnamefont
  {Streets}},\ }\emph {\bibinfo {title} {{Ricci Yang-Mills Flow}}},\ \href@noop
  {} {Ph.D. thesis},\ \bibinfo  {school} {Duke U. (main)} (\bibinfo {year}
  {2007})\BibitemShut {NoStop}%
\bibitem [{\citenamefont {Topping}(2006)}]{topping_2006}%
  \BibitemOpen
  \bibfield  {author} {\bibinfo {author} {\bibfnamefont {P.}~\bibnamefont
  {Topping}},\ }\href {\doibase 10.1017/CBO9780511721465} {\emph {\bibinfo
  {title} {Lectures on the Ricci Flow}}},\ London Mathematical Society Lecture
  Note Series\ (\bibinfo  {publisher} {Cambridge University Press},\ \bibinfo
  {year} {2006})\BibitemShut {NoStop}%
\bibitem [{\citenamefont {Pappas}(1984)}]{doi:10.1119/1.13934}%
  \BibitemOpen
  \bibfield  {author} {\bibinfo {author} {\bibfnamefont {R.~C.}\ \bibnamefont
  {Pappas}},\ }\href {\doibase 10.1119/1.13934} {\bibfield  {journal} {\bibinfo
   {journal} {American Journal of Physics}\ }\textbf {\bibinfo {volume} {52}},\
  \bibinfo {pages} {255} (\bibinfo {year} {1984})},\ \Eprint
  {http://arxiv.org/abs/https://doi.org/10.1119/1.13934}
  {https://doi.org/10.1119/1.13934} \BibitemShut {NoStop}%
\bibitem [{\citenamefont {Luna}\ \emph {et~al.}(2015)\citenamefont {Luna},
  \citenamefont {Monteiro}, \citenamefont {O'Connell},\ and\ \citenamefont
  {White}}]{Luna:2015paa}%
  \BibitemOpen
  \bibfield  {author} {\bibinfo {author} {\bibfnamefont {A.}~\bibnamefont
  {Luna}}, \bibinfo {author} {\bibfnamefont {R.}~\bibnamefont {Monteiro}},
  \bibinfo {author} {\bibfnamefont {D.}~\bibnamefont {O'Connell}}, \ and\
  \bibinfo {author} {\bibfnamefont {C.~D.}\ \bibnamefont {White}},\ }\href
  {\doibase 10.1016/j.physletb.2015.09.021} {\bibfield  {journal} {\bibinfo
  {journal} {Phys. Lett. B}\ }\textbf {\bibinfo {volume} {750}},\ \bibinfo
  {pages} {272} (\bibinfo {year} {2015})},\ \Eprint
  {http://arxiv.org/abs/1507.01869} {arXiv:1507.01869 [hep-th]} \BibitemShut
  {NoStop}%
\bibitem [{\citenamefont {Bah}\ \emph {et~al.}(2020)\citenamefont {Bah},
  \citenamefont {Dempsey},\ and\ \citenamefont {Weck}}]{Bah:2019sda}%
  \BibitemOpen
  \bibfield  {author} {\bibinfo {author} {\bibfnamefont {I.}~\bibnamefont
  {Bah}}, \bibinfo {author} {\bibfnamefont {R.}~\bibnamefont {Dempsey}}, \ and\
  \bibinfo {author} {\bibfnamefont {P.}~\bibnamefont {Weck}},\ }\href {\doibase
  10.1007/JHEP02(2020)180} {\bibfield  {journal} {\bibinfo  {journal} {JHEP}\
  }\textbf {\bibinfo {volume} {02}},\ \bibinfo {pages} {180} (\bibinfo {year}
  {2020})},\ \Eprint {http://arxiv.org/abs/1910.04197} {arXiv:1910.04197
  [hep-th]} \BibitemShut {NoStop}%
\bibitem [{\citenamefont {Carrillo-Gonz\'alez}\ \emph
  {et~al.}(2018)\citenamefont {Carrillo-Gonz\'alez}, \citenamefont {Penco},\
  and\ \citenamefont {Trodden}}]{Carrillo-Gonzalez:2017iyj}%
  \BibitemOpen
  \bibfield  {author} {\bibinfo {author} {\bibfnamefont {M.}~\bibnamefont
  {Carrillo-Gonz\'alez}}, \bibinfo {author} {\bibfnamefont {R.}~\bibnamefont
  {Penco}}, \ and\ \bibinfo {author} {\bibfnamefont {M.}~\bibnamefont
  {Trodden}},\ }\href {\doibase 10.1007/JHEP04(2018)028} {\bibfield  {journal}
  {\bibinfo  {journal} {JHEP}\ }\textbf {\bibinfo {volume} {04}},\ \bibinfo
  {pages} {028} (\bibinfo {year} {2018})},\ \Eprint
  {http://arxiv.org/abs/1711.01296} {arXiv:1711.01296 [hep-th]} \BibitemShut
  {NoStop}%
\bibitem [{\citenamefont {Kim}\ \emph {et~al.}(2020)\citenamefont {Kim},
  \citenamefont {Lee}, \citenamefont {Monteiro}, \citenamefont {Nicholson},\
  and\ \citenamefont {Peinador~Veiga}}]{Kim:2019jwm}%
  \BibitemOpen
  \bibfield  {author} {\bibinfo {author} {\bibfnamefont {K.}~\bibnamefont
  {Kim}}, \bibinfo {author} {\bibfnamefont {K.}~\bibnamefont {Lee}}, \bibinfo
  {author} {\bibfnamefont {R.}~\bibnamefont {Monteiro}}, \bibinfo {author}
  {\bibfnamefont {I.}~\bibnamefont {Nicholson}}, \ and\ \bibinfo {author}
  {\bibfnamefont {D.}~\bibnamefont {Peinador~Veiga}},\ }\href {\doibase
  10.1007/JHEP02(2020)046} {\bibfield  {journal} {\bibinfo  {journal} {JHEP}\
  }\textbf {\bibinfo {volume} {02}},\ \bibinfo {pages} {046} (\bibinfo {year}
  {2020})},\ \Eprint {http://arxiv.org/abs/1912.02177} {arXiv:1912.02177
  [hep-th]} \BibitemShut {NoStop}%
\bibitem [{\citenamefont {Green}\ \emph {et~al.}(2012)\citenamefont {Green},
  \citenamefont {Schwarz},\ and\ \citenamefont {Witten}}]{Green:2012oqa}%
  \BibitemOpen
  \bibfield  {author} {\bibinfo {author} {\bibfnamefont {M.~B.}\ \bibnamefont
  {Green}}, \bibinfo {author} {\bibfnamefont {J.~H.}\ \bibnamefont {Schwarz}},
  \ and\ \bibinfo {author} {\bibfnamefont {E.}~\bibnamefont {Witten}},\ }\href
  {\doibase 10.1017/CBO9781139248563} {\emph {\bibinfo {title} {{Superstring
  Theory Vol. 1}: {25th Anniversary Edition}}}},\ Cambridge Monographs on
  Mathematical Physics\ (\bibinfo  {publisher} {Cambridge University Press},\
  \bibinfo {year} {2012})\BibitemShut {NoStop}%
\bibitem [{\citenamefont {Abouelsaood}\ \emph {et~al.}(1987)\citenamefont
  {Abouelsaood}, \citenamefont {Callan}, \citenamefont {Nappi},\ and\
  \citenamefont {Yost}}]{Abouelsaood:1986gd}%
  \BibitemOpen
  \bibfield  {author} {\bibinfo {author} {\bibfnamefont {A.}~\bibnamefont
  {Abouelsaood}}, \bibinfo {author} {\bibfnamefont {C.~G.}\ \bibnamefont
  {Callan}, \bibfnamefont {Jr.}}, \bibinfo {author} {\bibfnamefont {C.~R.}\
  \bibnamefont {Nappi}}, \ and\ \bibinfo {author} {\bibfnamefont {S.~A.}\
  \bibnamefont {Yost}},\ }\href {\doibase 10.1016/0550-3213(87)90164-7}
  {\bibfield  {journal} {\bibinfo  {journal} {Nucl. Phys. B}\ }\textbf
  {\bibinfo {volume} {280}},\ \bibinfo {pages} {599} (\bibinfo {year}
  {1987})}\BibitemShut {NoStop}%
\bibitem [{\citenamefont {Fradkin}\ and\ \citenamefont
  {Tseytlin}(1985)}]{Fradkin:1985qd}%
  \BibitemOpen
  \bibfield  {author} {\bibinfo {author} {\bibfnamefont {E.~S.}\ \bibnamefont
  {Fradkin}}\ and\ \bibinfo {author} {\bibfnamefont {A.~A.}\ \bibnamefont
  {Tseytlin}},\ }\href {\doibase 10.1016/0370-2693(85)90205-9} {\bibfield
  {journal} {\bibinfo  {journal} {Phys. Lett. B}\ }\textbf {\bibinfo {volume}
  {163}},\ \bibinfo {pages} {123} (\bibinfo {year} {1985})}\BibitemShut
  {NoStop}%
\bibitem [{\citenamefont {Callan}\ \emph {et~al.}(1987)\citenamefont {Callan},
  \citenamefont {Lovelace}, \citenamefont {Nappi},\ and\ \citenamefont
  {Yost}}]{Callan:1986bc}%
  \BibitemOpen
  \bibfield  {author} {\bibinfo {author} {\bibfnamefont {C.~G.}\ \bibnamefont
  {Callan}, \bibfnamefont {Jr.}}, \bibinfo {author} {\bibfnamefont
  {C.}~\bibnamefont {Lovelace}}, \bibinfo {author} {\bibfnamefont {C.~R.}\
  \bibnamefont {Nappi}}, \ and\ \bibinfo {author} {\bibfnamefont {S.~A.}\
  \bibnamefont {Yost}},\ }\href {\doibase 10.1016/0550-3213(87)90227-6}
  {\bibfield  {journal} {\bibinfo  {journal} {Nucl. Phys. B}\ }\textbf
  {\bibinfo {volume} {288}},\ \bibinfo {pages} {525} (\bibinfo {year}
  {1987})}\BibitemShut {NoStop}%
\bibitem [{\citenamefont {Kawai}\ \emph {et~al.}(1986)\citenamefont {Kawai},
  \citenamefont {Lewellen},\ and\ \citenamefont {Tye}}]{Kawai:1985xq}%
  \BibitemOpen
  \bibfield  {author} {\bibinfo {author} {\bibfnamefont {H.}~\bibnamefont
  {Kawai}}, \bibinfo {author} {\bibfnamefont {D.~C.}\ \bibnamefont {Lewellen}},
  \ and\ \bibinfo {author} {\bibfnamefont {S.~H.~H.}\ \bibnamefont {Tye}},\
  }\href {\doibase 10.1016/0550-3213(86)90362-7} {\bibfield  {journal}
  {\bibinfo  {journal} {Nucl. Phys. B}\ }\textbf {\bibinfo {volume} {269}},\
  \bibinfo {pages} {1} (\bibinfo {year} {1986})}\BibitemShut {NoStop}%
\bibitem [{\citenamefont {Polchinski}(2007)}]{Polchinski:1998rq}%
  \BibitemOpen
  \bibfield  {author} {\bibinfo {author} {\bibfnamefont {J.}~\bibnamefont
  {Polchinski}},\ }\href {\doibase 10.1017/CBO9780511816079} {\emph {\bibinfo
  {title} {{String theory. Vol. 1: An introduction to the bosonic string}}}},\
  Cambridge Monographs on Mathematical Physics\ (\bibinfo  {publisher}
  {Cambridge University Press},\ \bibinfo {year} {2007})\BibitemShut {NoStop}%
\bibitem [{\citenamefont {Pasarin}\ and\ \citenamefont
  {Tseytlin}(2020)}]{Pasarin:2020qoa}%
  \BibitemOpen
  \bibfield  {author} {\bibinfo {author} {\bibfnamefont {O.}~\bibnamefont
  {Pasarin}}\ and\ \bibinfo {author} {\bibfnamefont {A.~A.}\ \bibnamefont
  {Tseytlin}},\ }\href {\doibase 10.1016/j.physletb.2020.135594} {\bibfield
  {journal} {\bibinfo  {journal} {Phys. Lett. B}\ }\textbf {\bibinfo {volume}
  {807}},\ \bibinfo {pages} {135594} (\bibinfo {year} {2020})},\ \Eprint
  {http://arxiv.org/abs/2005.12396} {arXiv:2005.12396 [hep-th]} \BibitemShut
  {NoStop}%
\bibitem [{\citenamefont {Easson}\ \emph {et~al.}(2020)\citenamefont {Easson},
  \citenamefont {Keeler},\ and\ \citenamefont {Manton}}]{Easson:2020esh}%
  \BibitemOpen
  \bibfield  {author} {\bibinfo {author} {\bibfnamefont {D.~A.}\ \bibnamefont
  {Easson}}, \bibinfo {author} {\bibfnamefont {C.}~\bibnamefont {Keeler}}, \
  and\ \bibinfo {author} {\bibfnamefont {T.}~\bibnamefont {Manton}},\ }\href
  {\doibase 10.1103/PhysRevD.102.086015} {\bibfield  {journal} {\bibinfo
  {journal} {Phys. Rev. D}\ }\textbf {\bibinfo {volume} {102}},\ \bibinfo
  {pages} {086015} (\bibinfo {year} {2020})},\ \Eprint
  {http://arxiv.org/abs/2007.16186} {arXiv:2007.16186 [gr-qc]} \BibitemShut
  {NoStop}%
\end{thebibliography}%
\bibliographystyle{apsrev4-1}
\end{document}